\pdfoutput=1
\documentclass[sigconf]{acmart}

\usepackage{microtype}

\copyrightyear{2026}
\acmYear{2026}
\setcopyright{cc}
\setcctype{by}
\acmConference[UIST '26]{The 39th Annual ACM Symposium on User Interface Software and Technology}{November 02--05, 2026}{Detroit, MI, USA}
\acmBooktitle{The 39th Annual ACM Symposium on User Interface Software and Technology (UIST '26), November 02--05, 2026, Detroit, MI, USA}
\acmDOI{10.1145/3830398.3830673}
\acmISBN{979-8-4007-2856-3/2026/11}

\begin{document}

\title[EITWatch: Smartwatch-Integrated Planar EIT]{EITWatch: Smartwatch-Integrated Planar Electrical Impedance Tomography for Hand Gesture Recognition}

\author{Xuanyou Liu}
\authornote{Both authors contributed equally to this research. Xuanyou Liu is the corresponding author.}
\orcid{0009-0007-5326-6194}
\affiliation{%
  \department{Computer Science}
  \institution{Northwestern University}
  \city{Evanston}
  \state{Illinois}
  \country{USA}}
\email{xuanyou@u.northwestern.edu}

\author{Novel Alam}
\authornotemark[1]
\orcid{0009-0002-8720-4040}
\affiliation{%
  \department{Computer Science}
  \institution{Northwestern University}
  \city{Evanston}
  \state{Illinois}
  \country{USA}}
\email{novelalam2027@u.northwestern.edu}

\author{Karan Ahuja}
\orcid{0000-0003-2497-0530}
\affiliation{%
  \department{Computer Science}
  \institution{Northwestern University}
  \city{Evanston}
  \state{Illinois}
  \country{USA}}
\email{kahuja@northwestern.edu}

\renewcommand{\shortauthors}{Liu et al.}

\begin{abstract}
  Wrist Electrical Impedance Tomography (EIT) senses hand gestures from muscle- and tendon-driven impedance changes, but prior wrist-EIT systems require electrode coverage beyond the watch-back contact patch and separate analog front ends. We present EITWatch, the first wrist-EIT system built around smartwatch case-back geometry, asking whether this contact patch alone can support gesture recognition: eight planar electrodes in a 31\,mm ring acquire 35 impedance measurements at 48\,Hz. Because a planar array cannot encircle the wrist, EITWatch uses \emph{multi-depth scanning} to sample multiple source-sink distances and current paths; it beat matched adjacent injection by 15.1/10.4 percentage points (macro/micro) across all 12 participants. In a prompted study, within-session leave-one-round-out accuracy reached 91.4\%\slash 92.5\% (window\slash trial) for six macro-gestures, and 90.1\%\slash 91.5\% (window\slash segment) for five micro-gestures plus \emph{relax}; window-level cross-session and leave-one-user-out transfer reached 73.2\%\slash 70.4\% and 63.1\%\slash 55.3\% (macro\slash micro).
\end{abstract}



\maketitle

\begin{table*}[t]
  \caption{Hardware and task scope of electrical impedance sensing systems for hand gesture recognition. Circ.: circumferential; Local 2\texttimes2: a localized 4-electrode array carried on a wristband rather than encircling the wrist. Frame rates are as reported by each system and are not directly comparable, since several scale with electrode count.}
  \Description{A hardware comparison of seven electrical impedance sensing systems. The rows list electrode count, frame rate, electrode placement, form factor, and gesture task. Prior wrist systems use circumferential bands, a localized four-electrode array on a wristband, or a flexible bandage, while EITWatch uses eight planar dorsal electrodes on a watch back.}
  \label{tab:comparison}
  \centering
  \footnotesize
  {%
  \begin{tabular*}{\textwidth}{@{\extracolsep{\fill}}l ccccccc@{}}
    \toprule
    & Tomo~\cite{Tomo2015} & HighRes~\cite{HighResEIT2016} & EIT-kit~\cite{EITKit2021} & EITPose~\cite{EITPose2024} & EI-Lite~\cite{EILite2025} & BandEI~\cite{BandEI2025} & \textbf{EITWatch} \\
    \midrule
    Electrodes          & 8            & 8 / 16 / 32  & Up to 64      & 8          & 4          & 32             & 8 \\
    Frame rate (Hz)     & 10           & 3--100       & $\leq$10      & 10         & 100        & 30             & 48 \\
    Electrode placement & Circ.        & Circ.        & Circ.         & Circ.      & Local 2\texttimes2 & Finger + wrist & Planar (dorsal) \\
    Form factor         & Forearm band & Forearm band & Configurable  & Wrist band & Wrist band & Flex.\ bandage & Watch back \\
    Gesture task        & Macro        & Macro        & Configurable  & Macro      & Micro      & Macro + micro  & Macro + micro \\
    \bottomrule
  \end{tabular*}%
  }
\end{table*}

\section{Introduction}

Smartwatches sit on the wrist all day, maintaining continuous skin contact through the case back, a contact that has rarely been used as an interactive surface. Electrical Impedance Tomography (EIT) is well suited to exploit this contact: by injecting small alternating currents and measuring the resulting voltage patterns, it can detect impedance shifts caused by muscle and tendon deformation beneath the skin~\cite{Tomo2015,EITPose2024}. Prior wrist-EIT systems show accurate gesture recognition, but they are worn as separate accessories, either circumferential bands~\cite{Tomo2015,EITKit2021,EITPose2024} or a dedicated wristband carrying a localized array~\cite{EILite2025}, and connected to external analog front ends. This leaves a direct HCI question unanswered: can the watch-back contact patch alone serve as a viable EIT sensing site for gesture recognition?

\begin{figure}[t]
  \centering
  \includegraphics[width=\columnwidth]{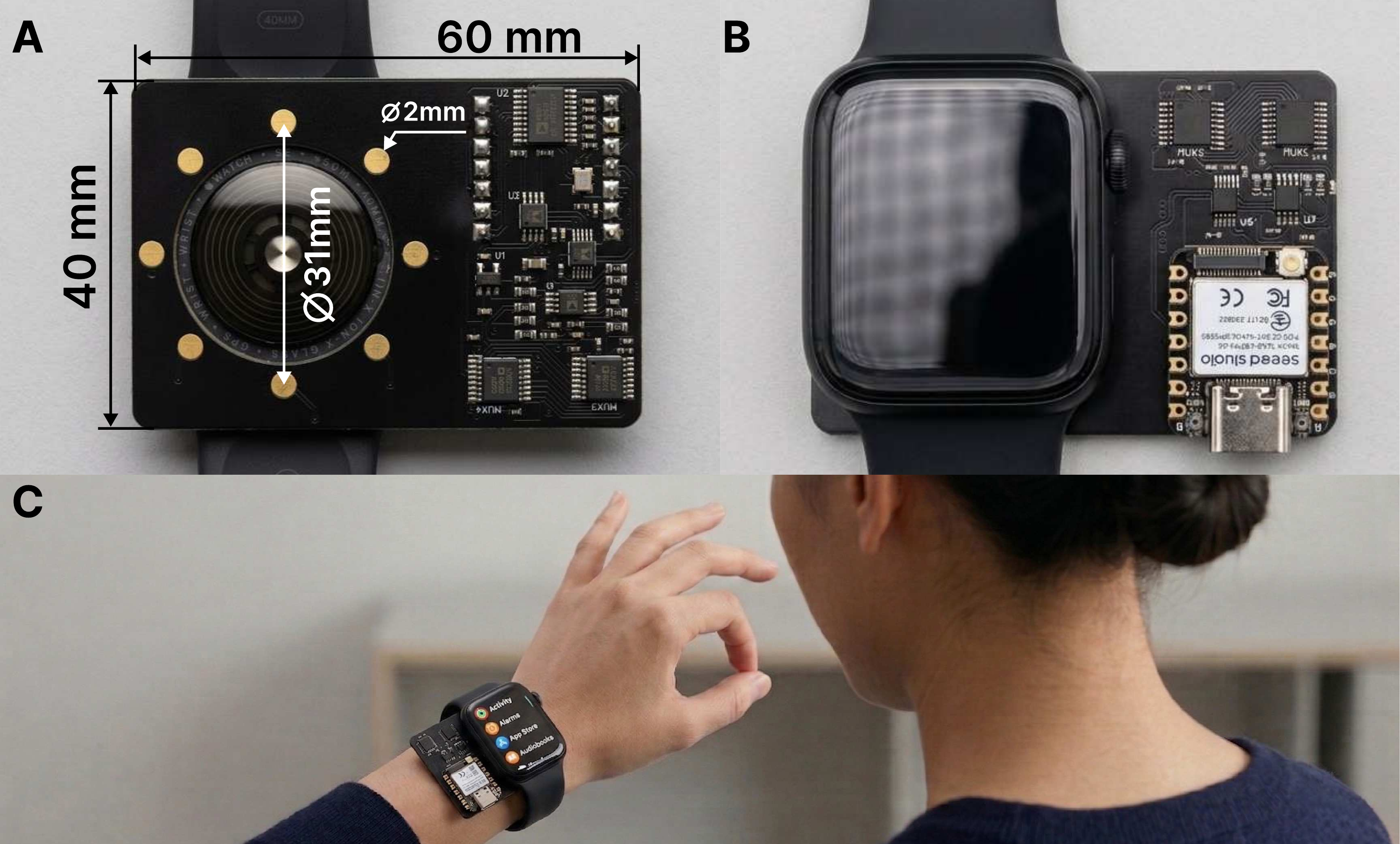}
  \caption{EITWatch overview. (A) The custom PCB with eight planar EIT electrodes arranged in a 31\,mm diameter ring on a 40\,$\times$\,60\,mm board. (B) The board mounted beneath a standard 40\,mm watch case. (C) The system worn on the wrist during gesture input.}
  \Description{Three panels show the EITWatch prototype. The first shows a 40 by 60 mm circuit board with eight circular electrodes in a 31 mm ring, the second shows the board mounted under a 40 mm watch case, and the third shows the assembled prototype on the dorsal wrist.}
  \label{fig:overview}
\end{figure}

We present \textbf{EITWatch}, the first wrist-EIT system centered on smartwatch case-back geometry (Figure~\ref{fig:overview}). The prototype features a custom PCB with eight skin-facing planar electrodes arranged in a 31\,mm ring; an onboard ESP32-S3 microcontroller sequences these electrodes to acquire 35 impedance measurements at a frame rate of 48\,Hz. The current discrete analog front end extends 20\,mm beyond the case, so the contribution is not a fully miniaturized smartwatch; rather, it isolates the watch-back sensing geometry and tests whether that contact area alone provides discriminative impedance measurements.

Moving to a planar watch-back geometry also changes the acquisition design. A dorsal array cannot surround the wrist cross-section, so the adjacent-injection schedule used by circumferential systems mainly emphasizes shallow tissue. EITWatch therefore uses \emph{multi-depth scanning}, a multi-distance, multi-path protocol that sweeps source-sink separation and direction within each frame. In a homogeneous half-space, its source-sink distances correspond to heuristic characteristic depth scales from approximately 6\,mm to 15.5\,mm (Section~\ref{sec:theory}); on the heterogeneous wrist, changing the sink also changes the angular path through muscles and tendons. We validate the protocol in a water tank and through a matched on-body ablation.

Across 12 participants, within-session leave-one-round-out evaluation exceeded 90\% window-level accuracy for both gesture sets, measuring within-session repeatability rather than deployable recognition; 48-hour cross-session and leave-one-user-out transfer dropped substantially, and a matched ablation favored multi-depth scanning over adjacent injection on the same hardware and tasks (Section~\ref{sec:results}). Together, these results indicate that watch-back EIT carries discriminative signal, while re-donning and person-independent transfer remain open challenges.

The contributions of this paper are (1)~\textbf{the EITWatch device concept}: the first wrist-EIT system built around a standard smartwatch case-back contact area, isolating the sensing geometry to the watch-back contact patch even though the current electronics are not yet fully miniaturized; and (2)~\textbf{feasibility evidence}: a 12-participant evaluation of within-session, cross-session, and cross-user recognition, a matched comparison showing that multi-depth scanning outperforms adjacent injection on this planar system, and measurements of prototype latency and power.

To support future research in planar wrist-EIT, we release the EITWatch schematics, PCB layouts, and firmware at \url{https://github.com/XuanyouLiu/EITWatch_Hardware}.

\section{Related Work}

\paragraph{Wrist-worn EIT}
Prior wrist-EIT systems show that the wrist exhibits rich impedance structure for gesture recognition, but all require hardware beyond the watch-back contact patch. Tomo~\cite{Tomo2015}, HighRes~\cite{HighResEIT2016}, and EIT-kit~\cite{EITKit2021} use circumferential forearm or wrist bands; EITPose~\cite{EITPose2024} and EI-Lite~\cite{EILite2025} reduce electrode count but still rely on a separate band. In each case, the electrode array connects to an external analog front end board that often remains larger than the wearable band alone. BandEI~\cite{BandEI2025} explores flexible placement including the wrist, yet performs best with 24 finger channels rather than 8 wrist-only channels. EITWatch instead places all electrodes on the case back, asking whether a watch-back-sized planar array can support gesture recognition at all. Table~\ref{tab:comparison} situates hardware and task scope only; the paper's main evidence comes from matched experiments on the same device, especially the protocol ablation in Section~\ref{sec:ablation_cross}.

\paragraph{Smartwatch gesture input}
Non-EIT smartwatch input includes EMG~\cite{Saponas2009}, wrist gestures~\cite{Gong2016}, bio-acoustics~\cite{Laput2016}, ultrasonic beamforming~\cite{Iravantchi2019}, active sonar~\cite{FingerIO2016}, body-coupled RF~\cite{SkinTrack2016}, and depth sensing~\cite{Sridhar2017}. Most require additional hardware or active signal emission beyond the watch itself; EITWatch instead asks whether the existing case-back contact patch can become a sensing surface in its own right.

\paragraph{Surface planar EIT}
Planar EIT places all electrodes on one surface and senses into the volume beneath it. Prior work established reconstruction algorithms~\cite{Mueller1999}, demonstrated probe-based planar sensing at depths on the order of 2--4\,cm in controlled phantoms~\cite{Kao2006}, and characterized depth-dependent resolution for one-sided geometries~\cite{ChenSoleimani2019}. That literature makes watch-back sensing physically plausible, but does not address smartwatch-scale planar arrays for gesture recognition. EITWatch addresses that device-level question directly.

\begin{figure*}[!t]
  \centering
  \includegraphics[width=\textwidth]{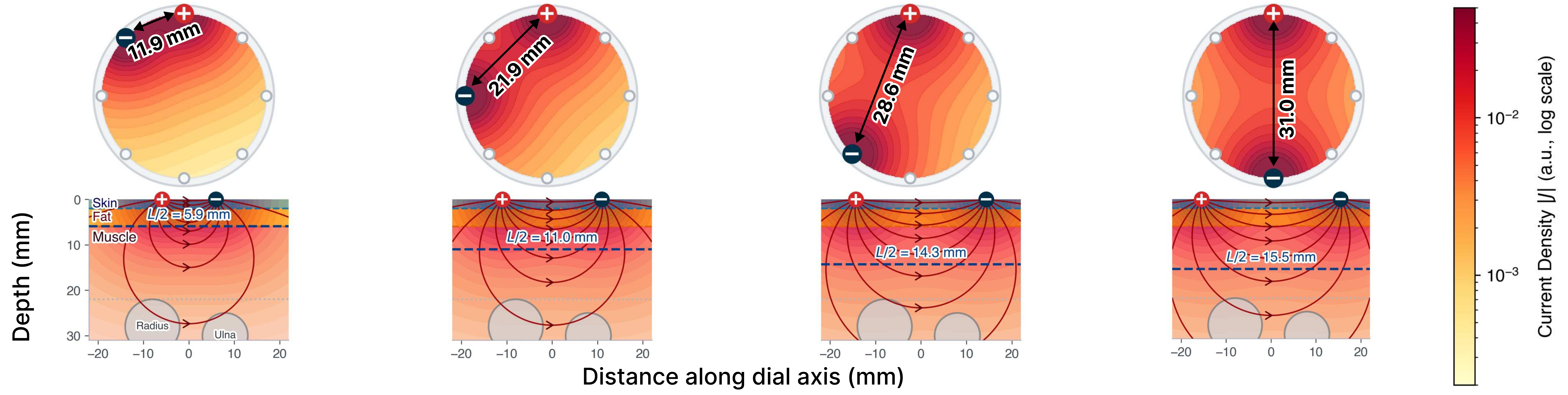}
  \caption{Current penetration in a homogeneous half-space. Varying the injection-pair distance $L$ from adjacent to opposite increases the heuristic characteristic depth scale ($L/2$); current density nevertheless decays continuously with depth. }
  \Description{Four cross-sectional simulations show current-density fields beneath a planar electrode array as electrode separation increases from 11.9 to 31.0 mm. Wider separation extends the field progressively deeper below the surface.}
  \label{fig:theory_ops}
\end{figure*}

\section{Planar Sensing on the Watch Back}
\label{sec:theory}

EITWatch places all eight electrodes on the dorsal wrist surface within a 31\,mm ring. To understand what this compact planar array can sense, we examine how current penetration varies with electrode spacing.

When a small alternating current is injected between two surface electrodes separated by distance $L$, the resulting electric potential $\phi$ inside the tissue domain $\Omega$ satisfies the quasi-static Laplace equation~\cite{Bayford2006,Kao2006}:
\begin{equation}
    \nabla \cdot (\sigma \nabla \phi) = 0 \quad \text{in } \Omega
\end{equation}
where $\sigma$ is the spatial conductivity distribution. The current density $\mathbf{J} = -\sigma \nabla \phi$ does not flow in a straight line between the electrodes; instead, it curves downward into the tissue volume.
For a homogeneous half-space, the current density magnitude along the depth axis $z$ is:
\begin{equation}
    |\mathbf{J}(z)| \propto \frac{1}{\bigl((L/2)^2 + z^2\bigr)^{3/2}}
\end{equation}
This solution shows that current density decays continuously with depth. We use $L/2$ as a heuristic characteristic depth scale, rather than a discrete penetration boundary~\cite{Grimnes2008}. As shown in Figure~\ref{fig:theory_ops}, adjacent electrodes ($L \approx 11.9$\,mm) weight shallower regions more strongly, whereas diametrically opposite electrodes ($L \approx 31.0$\,mm) extend the field farther into the half-space. This homogeneous model isolates the effect of separation; it does not model wrist anatomy.

Gesture recognition exploits the impedance modulation produced as muscle contracts and its fibers shorten and thicken. Surface measurements on the anterior forearm at 50\,kHz, EITWatch's operating frequency, report that resistance \emph{increases} under isometric finger-flexor contraction~\cite{Shiffman2003}, and localized impedance myography of the biceps places the magnitude near 10\% at maximal voluntary contraction~\cite{Li2016}. This modulation acts against a static background of poorly conducting tissue, including cortical bone ($\sim$0.02\,S/m) and fat ($\sim$0.02--0.04\,S/m)~\cite{Gabriel1996c}. Because EITWatch senses from the dorsal surface only, bone and fat steer the field and create position-dependent sensitivity rather than the more uniform wrap-around view of circumferential systems. Multi-depth scanning varies both separation and sink direction, thereby sampling multiple depth profiles and angular paths within each frame. The water-tank study below compares the complete protocols under controlled conditions, while the on-body ablation evaluates them on real wrist anatomy; neither comparison isolates separation from angular coverage or channel count.

\section{EITWatch System Design}
\label{sec:system_design}

\subsection{Electrode Array}

Eight gold-plated stainless steel disk electrodes (diameter 2\,mm) are mounted on the skin-facing surface of a custom PCB (Figure~\ref{fig:overview}A). They are evenly distributed on a $\varnothing$\,31\,mm circle, yielding $\sim$11.9\,mm adjacent spacing. Because the ring fits within the 40\,mm watch case back, the board attaches directly to the case and presses all electrodes flush against the dorsal wrist.

\subsection{PCB and Firmware}
\label{sec:system_design:pcb}

\begin{figure}[t]
  \centering
  \includegraphics[width=\columnwidth]{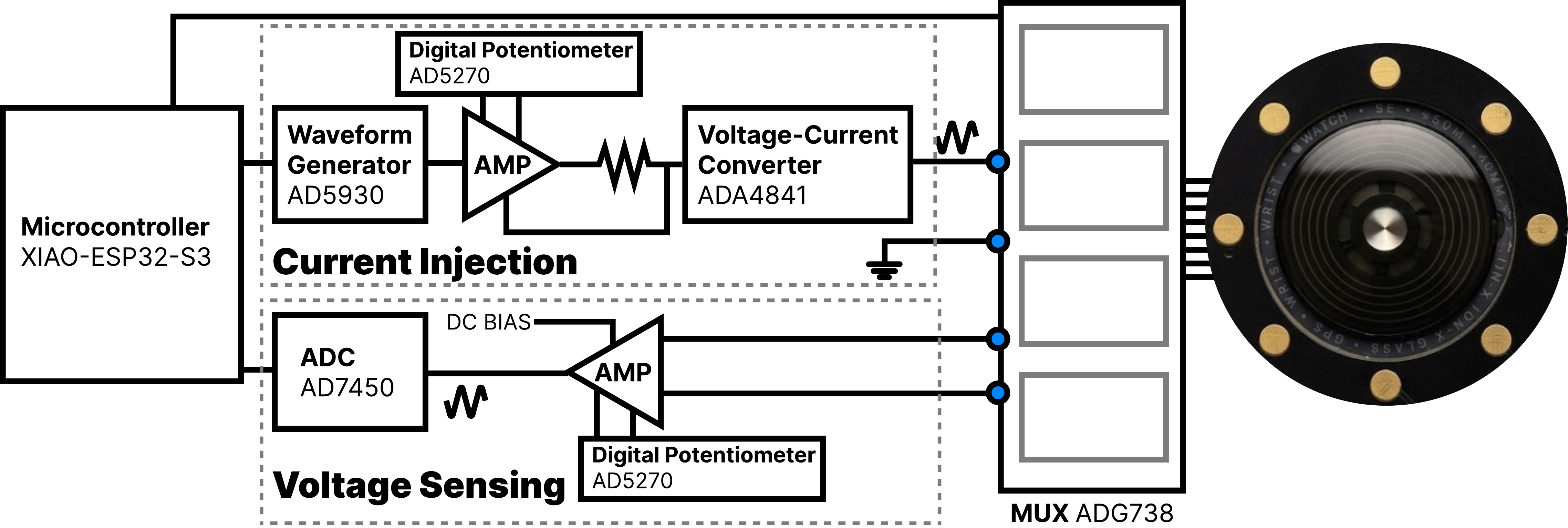}
  \caption{EITWatch circuit architecture. An ESP32-S3 microcontroller sequences eight electrodes through a 35-channel multi-depth scanning schedule at $\sim$48\,Hz.}
  \Description{A block diagram shows an ESP32-S3 controlling a 50 kHz waveform generator, current source, instrumentation amplifier, analog-to-digital converter, digital potentiometers, and switch matrices connected to eight electrodes.}
  \label{fig:hardware}
\end{figure}

Figure~\ref{fig:hardware} shows the circuit architecture. The custom six-layer PCB (40\,$\times$\,60\,mm) carries eight gold electrodes on the bottom and mixed-signal conditioning on top. The electrode ring and switch matrices occupy the 40\,mm region beneath the watch case, while the remaining 20\,mm extension houses the discrete analog front end. This partition matches the watch-back footprint while keeping the prototype debuggable with off-the-shelf components before full electronics miniaturization.
An AD5930~\cite{AD5930} waveform generator produces a 50\,kHz sinusoid that is amplified and converted to a current source by an ADA4841-based voltage-to-current stage. An AD5270 digital potentiometer sets the injected current to approximately 1\,mA peak-to-peak. On the sensing side, an AD8220 instrumentation amplifier with AD5270-adjustable gain amplifies the differential voltage, which is digitized by a 12-bit ADC (AD7450~\cite{AD7450}).
Four 8-channel matrix switches (ADG738) route injection and sense pairs to the electrodes for the multi-depth schedule (Section~\ref{sec:measurement_config}).

A Seeed XIAO ESP32-S3 microcontroller running FreeRTOS sequences the scanning schedule at 48\,Hz, chosen to resolve the sub-second transients that characterize micro-gestures. For the gesture recognition study (Section~\ref{sec:study}), the device streams the 35 scalar features over Wi-Fi to a host computer for logging and offline classification.

\subsection{Measurement Configuration}
\label{sec:measurement_config}

As a baseline we use four-pole adjacent injection, the standard in circumferential wrist EIT: the source-sink pair occupies neighboring electrodes and steps around the ring, with voltage sensed differentially on non-drive electrodes. With 8 electrodes and 5 non-drive differential measurements per drive pair, this yields a 40-channel baseline used in Section~\ref{sec:ablation_cross}. On a dorsal planar array, however, this protocol is shallow, concentrating current near the skin surface ($z \approx L/2 \approx 6$\,mm) and reducing sensitivity to deeper muscle deformation.

We therefore use a \emph{multi-depth scanning} protocol. The electrode at the 12-o'clock position serves as the fixed positive source; the negative sink iterates through the remaining 7 positions, producing injection pairs from adjacent ($L \approx 11.9$\,mm) to diametrically opposite ($L \approx 31.0$\,mm). Five four-pole differential measurements per injection pair yield $7 \times 5 = 35$ measurements per frame, so only the injection geometry differs from the baseline. The schedule samples multiple source-sink distances and angular paths in one cycle. A pilot comparison across all eight possible source electrodes on two participants showed less than 1.5 percentage points of variation in LORO accuracy, indicating that source choice does not meaningfully affect recognition on this symmetric ring layout. This schedule trades a small reduction in per-frame channel count for broader multi-depth, multi-path coverage.

\subsection{Water-Tank Validation}

\begin{figure}[tb]
  \centering
  \includegraphics[width=\columnwidth]{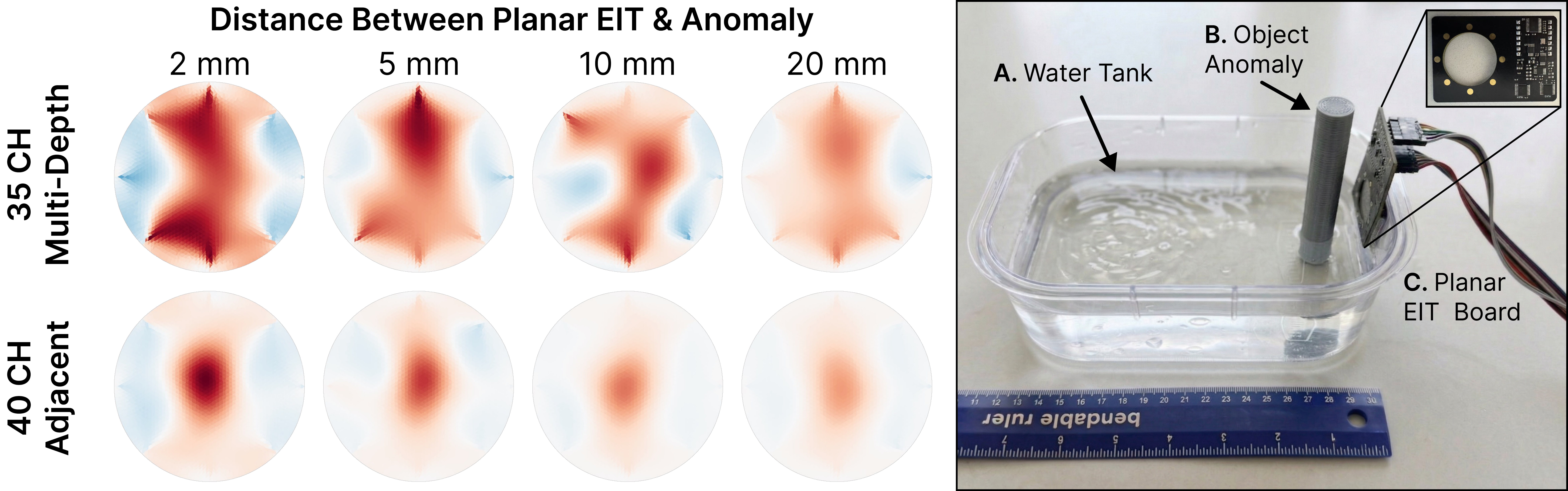}
  \caption{Water-tank validation across four anomaly depths (2, 5, 10, and 20\,mm). Multi-depth scanning retains stronger deep-anomaly contrast than adjacent injection at 10--20\,mm, consistent with greater deep-anomaly sensitivity but not isolating depth as the cause.}
  \Description{Photographs show a saline tank, conductive target, and planar EIT board. Reconstructed impedance images compare 35-channel multi-depth scanning with a 40-channel adjacent-injection baseline at target depths of 2, 5, 10, and 20 mm. Multi-depth contrast remains more visible at 10 and 20 mm.}
  \label{fig:watertank}
\end{figure}

We compared the complete protocols in a saline tank with a conductive anomaly at 2, 5, 10, and 20\,mm from the electrode surface (Figure~\ref{fig:watertank}), using identical hardware and 10 repeats per depth. For each trial, we reconstructed the differential signal with pyEIT~\cite{pyEIT} and quantified it as the mean absolute reconstructed value within a fixed ROI centered on the anomaly location. For each depth, responses were averaged over repeats and normalized by the corresponding 2\,mm value within the same protocol. Both protocols detected shallow anomalies similarly, with normalized response near 1.00 at 2\,mm and about 0.80 at 5\,mm. At greater depths, however, the multi-depth protocol retained a larger normalized response: at 10\,mm, 0.58\,$\pm$\,0.04 versus 0.37\,$\pm$\,0.05 for adjacent injection, and at 20\,mm, 0.31\,$\pm$\,0.03 versus 0.14\,$\pm$\,0.02. This result is consistent with stronger sensitivity to deeper anomalies under multi-depth scanning. However, because the protocols also differ in source-sink separation, angular current paths, and per-frame channel count (35 vs.\ 40), it does not isolate depth as the cause of the difference.

\begin{table*}[t]
  \caption{Per-participant recognition accuracy (\%). Macro/Micro LORO use within-session multi-depth scanning; Trial and Segment are majority votes over a prompted hold or an annotated interval. Cross-session, LOUO, and Adjacent are \emph{window-level}; Adjacent is matched LORO under 40-channel adjacent injection. Mean $\pm$: across-participant sample SD. Round SD: mean of per-participant fold SDs.}
  \Description{A twelve-participant results table. Within-session macro columns report window-level and trial-level accuracy, and micro columns report window-level and segment-level accuracy. Cross-session, leave-one-user-out, and adjacent-injection columns are window-level only. The macro trial-level mean is 92.5 percent and the micro segment-level mean is 91.5 percent, matching the pooled confusion matrices.}
  \label{tab:results}
  \centering
  \footnotesize
  {%
  \begin{tabular*}{\textwidth}{@{\extracolsep{\fill}}l *{12}{c}@{}}
    \toprule
    & \multicolumn{3}{c}{Macro LORO} & \multicolumn{3}{c}{Micro LORO} & \multicolumn{2}{c}{Cross-session} & \multicolumn{2}{c}{LOUO} & \multicolumn{2}{c}{Adjacent} \\
    & \multicolumn{3}{c}{(multi-depth)} & \multicolumn{3}{c}{(multi-depth)} & \multicolumn{2}{c}{(window)} & \multicolumn{2}{c}{(window)} & \multicolumn{2}{c}{(window)} \\
    \cmidrule(lr){2-4} \cmidrule(lr){5-7} \cmidrule(lr){8-9} \cmidrule(lr){10-11} \cmidrule(lr){12-13}
    ID & Window & Trial & Round SD & Window & Segment & Round SD & Mac & Mic & Mac & Mic & Mac & Mic \\
    \midrule
    P1  & 93.0 & 94.0 & 10.3 & 91.4 & 94.5 & 3.3 & 67.6 & 78.0 & 64.9 & 58.8 & 78.4 & 85.4 \\
    P2  & 91.2 & 93.0 &  8.6 & 89.6 & 91.1 & 6.1 & 83.1 & 59.9 & 69.3 & 47.6 & 72.0 & 82.9 \\
    P3  & 95.8 & 95.7 &  6.0 & 93.6 & 95.4 & 2.8 & 78.2 & 63.6 & 59.6 & 40.8 & 77.0 & 87.2 \\
    P4  & 88.9 & 90.6 & 12.6 & 88.6 & 89.2 & 8.2 & 72.6 & 71.5 & 59.9 & 68.7 & 76.2 & 79.8 \\
    P5  & 86.3 & 87.9 & 16.6 & 87.7 & 85.9 & 7.5 & 64.9 & 61.7 & 59.8 & 46.7 & 72.9 & 71.7 \\
    P6  & 92.5 & 93.6 &  9.0 & 90.2 & 92.9 & 5.5 & 74.5 & 77.9 & 71.2 & 62.8 & 79.9 & 84.9 \\
    P7  & 89.4 & 91.0 & 10.9 & 89.2 & 90.0 & 6.3 & 84.3 & 79.4 & 54.7 & 59.8 & 72.9 & 75.0 \\
    P8  & 89.6 & 91.4 & 12.2 & 89.3 & 91.0 & 6.8 & 71.3 & 66.5 & 48.4 & 50.1 & 75.3 & 73.9 \\
    P9  & 94.8 & 94.4 &  6.0 & 91.5 & 94.2 & 2.8 & 79.8 & 77.0 & 76.4 & 63.5 & 77.7 & 81.8 \\
    P10 & 92.0 & 93.1 & 12.4 & 90.2 & 92.0 & 5.7 & 69.9 & 76.8 & 62.5 & 54.8 & 84.0 & 81.7 \\
    P11 & 95.6 & 95.1 &  6.1 & 91.7 & 93.4 & 2.8 & 65.0 & 64.1 & 63.7 & 52.1 & 75.7 & 79.1 \\
    P12 & 87.7 & 90.2 & 14.1 & 88.2 & 88.4 & 9.4 & 67.2 & 68.4 & 66.8 & 57.9 & 73.6 & 73.0 \\
    \midrule
    Mean & 91.4\,$\pm$\,3.1 & 92.5\,$\pm$\,2.3 & 10.4 & 90.1\,$\pm$\,1.7 & 91.5\,$\pm$\,2.8 & 5.6 & 73.2\,$\pm$\,6.8 & 70.4\,$\pm$\,7.2 & 63.1\,$\pm$\,7.5 & 55.3\,$\pm$\,8.1 & 76.3\,$\pm$\,3.4 & 79.7\,$\pm$\,5.2 \\
    \bottomrule
  \end{tabular*}%
  }
\end{table*}

\section{Gesture Recognition Study}
\label{sec:study}

\subsection{Gesture Set Design}
\label{sec:study_design}

\begin{figure}[!b]
  \centering
  \includegraphics[width=0.94\columnwidth]{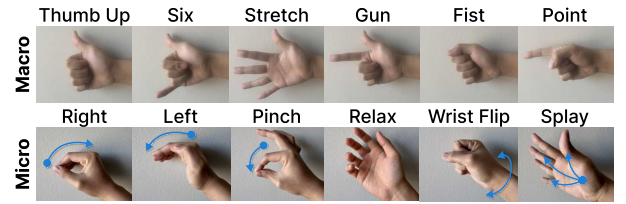}
  \caption{Gesture set. Top row: six static macro-gestures. Bottom row: six dynamic micro classes (including \emph{relax}); arrows indicate motion direction.}
  \Description{Two rows of hand-pose photographs. The top row shows six macro-gestures: Thumb Up, Six, Stretch, Gun, Fist, and Point. The bottom row shows six micro classes with motion arrows: Swipe Right, Swipe Left, Pinch, Relax, Wrist Flip, and Splay.}
  \label{fig:gestures}
\end{figure}

We evaluated two gesture vocabularies (Figure~\ref{fig:gestures}).

\paragraph{Macro-Gestures}
Six macro-gestures were used: \textbf{Six}, \textbf{Gun}, \textbf{Point}, \textbf{Thumb Up}, \textbf{Stretch}, and \textbf{Fist}. Three of these (Thumb Up, Stretch, and Fist) are drawn from the gesture set of Tomo~\cite{Tomo2015}.

\paragraph{Micro-Gestures}
Five micro-gestures were selected as subtle, eyes-free smartwatch commands: \textbf{Swipe Left} and \textbf{Swipe Right} (thumb slides across the index finger), \textbf{Pinch} (two rapid thumb-to-index pinches), \textbf{Splay} (rapid full-hand opening), and \textbf{Wrist Flip} (rapid forearm rotation $\approx$90\textdegree). A \emph{relax} class from all untagged intervals yields a six-class task on prompted recordings, letting us evaluate both gesture discrimination and relax-state rejection without claiming open-world online spotting.

\subsection{Participants and Apparatus}

Twelve right-handed participants (6F, 6M; age 20--31; wrist circumference 181\,$\pm$\,10.2\,mm; BMI 25.3\,$\pm$\,4.8\,kg/m$^2$) were recruited from the university community (IRB-approved; informed consent obtained). EITWatch was positioned on the distal dorsal wrist above the ulnar styloid and secured with a silicone strap without conductive gel; participants tightened it until the watch no longer slid during use. We did not perform additional contact-quality, pressure, or per-user firmware calibration so that performance would reflect the fixed watch-back geometry and acquisition protocol rather than user-specific compensation. A Meta Quest~3 provided VR prompts and optical hand-tracking verification.

\subsection{Procedure}

Each participant completed a $\sim$30\,min session in two blocks. Before recording, participants practiced with VR prompts and hand-tracking feedback. In the \emph{macro-gesture block}, 10 rounds of 6 gestures were presented in randomized order; the Quest~3 verified the pose, each gesture was held for 2\,s while EIT data were recorded, and a 10\,s break separated rounds. Because every macro trial was a prompted held pose, this closed-set block contained no idle interval to label as \emph{relax}. In the \emph{micro-gesture block}, 10 rounds of the 5 prompted transient gestures were collected, with 5 consecutive repetitions per round. Participants self-tagged onset/offset via spacebar, and a 5\,s break separated rounds. All untagged intervals, including inter-repetition and inter-round periods, were labeled \emph{relax}, which forms the sixth class and constitutes 51.7\% of micro-gesture windows. Separately, all participants completed a new study under the adjacent-injection protocol for the matched acquisition-protocol comparison. All participants returned 48 hours later for a second session after independently re-donning the device.

\subsection{Feature Extraction and Classification}

Raw 35-channel EIT frames were low-pass filtered with a zero-phase 4th-order Butterworth filter (0.5\,Hz macro / 1.0\,Hz micro cutoff) over each continuous recording before the LORO split, then IQR filtered ($k{=}3$) for outliers. The lower macro cutoff retains slowly varying held poses, whereas the higher micro cutoff retains faster transient motion. Because the inter-round breaks exceed the filters' effective span, this pre-split filtering cannot carry gesture information across a LORO boundary, and no windows overlap between training and held-out rounds. Macro-gestures used sliding windows of 40 frames with step size 2 ($\sim$0.83\,s at 48\,Hz); micro-gestures used 3 frames with step size 1 ($\sim$0.06\,s). From each window we computed five statistics per channel (mean, SD, slope, range, and net displacement), giving 175 features per window ($35 \times 5$). An ExtraTrees classifier (200 estimators macro, 400 micro; balanced class weighting) was used. IQR filtering, feature extraction, and model selection were performed independently within each fold; window, cutoff, and tree-count settings were fixed using a validation round from the training data before the test round was scored. LORO is therefore a user-dependent within-session repeatability metric, not a deployment estimate. We report accuracy at three granularities: \emph{window} (each classifier decision), \emph{trial} (majority vote over windows in a prompted macro hold), and \emph{segment} (majority vote over windows in an annotated micro-gesture or contiguous relax interval). Window-level accuracy is the primary per-decision metric because trial and segment voting assume known interval boundaries. This study evaluates classification on prompted recordings rather than open-world onset detection.

\section{Results}
\label{sec:results}

We evaluate EITWatch along three axes: within-session discriminability, robustness across sessions and users, and protocol choice, then report the on-device latency and power cost. Table~\ref{tab:results} reports per-participant accuracies and foregrounds the distinction between user-dependent within-session results and the harder transfer settings.

\subsection{Within-Session Gesture Recognition}

\begin{figure}[t]
  \centering
  \includegraphics[width=\columnwidth]{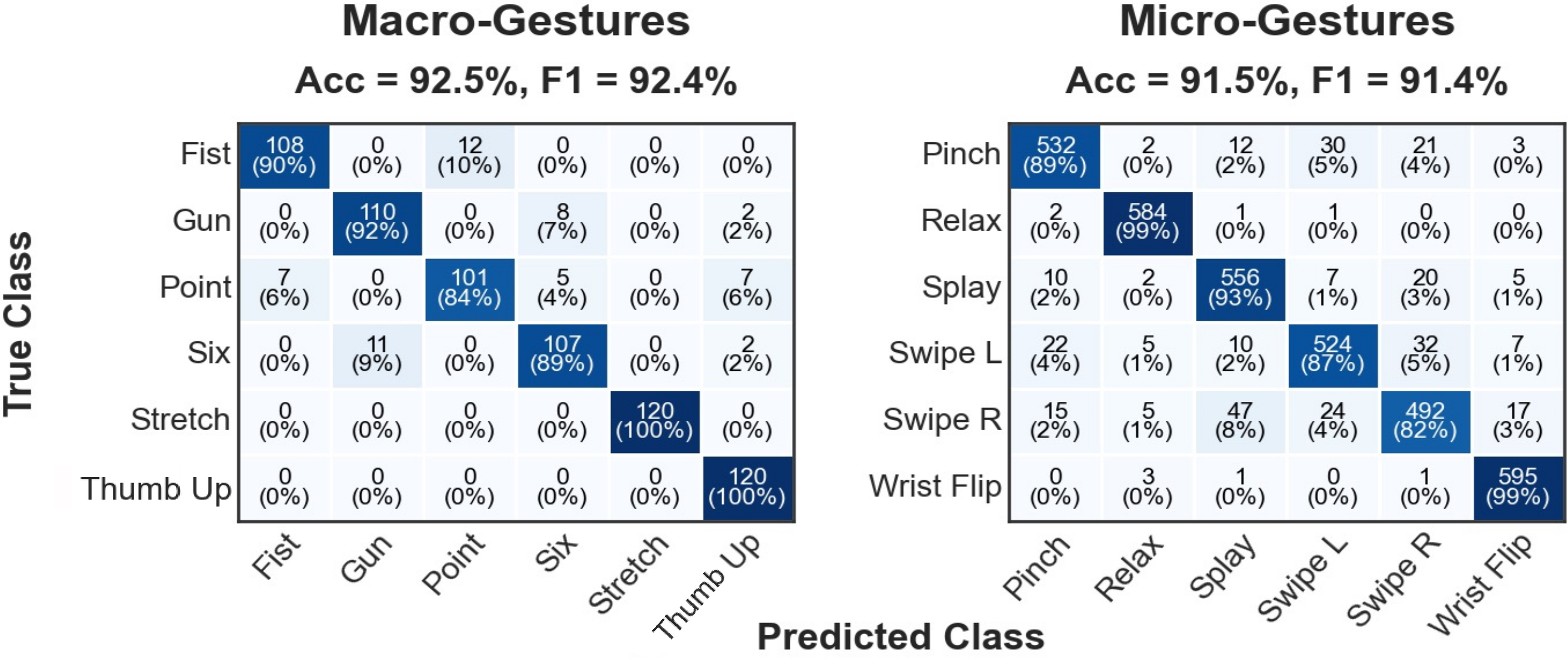}
  \caption{Pooled leave-one-round-out confusion matrices ($N=12$). Left: macro-gestures at the \emph{trial level}. Right: micro-gestures at the \emph{segment level}. Cells report counts and row-normalized percentages.}
  \Description{Two side-by-side leave-one-round-out confusion matrices across 12 participants. The left matrix shows macro trial-level recognition with 92.5 percent accuracy and 92.4 percent F1; Stretch and Thumb Up are perfect, while Point is most often confused with Fist or Thumb Up. The right matrix shows micro segment-level recognition, including relax, with 91.5 percent accuracy and 91.4 percent F1; Relax and Wrist Flip reach 99 percent, while Swipe Right is most often confused with Splay and Swipe Left.}
  \label{fig:confusion}
\end{figure}

For macro-gestures, mean LORO accuracy was 91.4\% (SD\,=\,3.1) per window and 92.5\% (SD\,=\,2.3) after trial voting, with a mean per-round SD of 10.4 across leave-one-round folds. For the prompted six-class micro task, it was 90.1\% (SD\,=\,1.7) per window and 91.5\% (SD\,=\,2.8) after segment-level voting within annotated segments, with a mean per-round SD of 5.6. Figure~\ref{fig:confusion} shows the pooled macro trial-level and micro segment-level confusion matrices across all 12 participants (macro Acc\,=\,92.5\%, F1\,=\,92.4\%; micro Acc\,=\,91.5\%, F1\,=\,91.4\%). Stretch and Thumb Up were recognized perfectly; Point was the weakest macro class (84\%), mainly confused with Fist or Thumb Up. On the micro side, Relax and Wrist Flip reached 99\%, while Swipe Right was hardest (82\%), mainly confused with Splay and Swipe Left. These within-session results measure repeatability within a single donning session; they do not establish person-independent or open-world performance. The micro \emph{relax} class accounts for 51.7\% of windows; balanced class weights and the primary window-level metric limit the influence of this imbalance.

\subsection{Cross-Session and Cross-User Robustness}

All 12 participants returned after 48 hours and re-donned EITWatch unassisted. For each participant, we trained a model on all first-session data and evaluated it on that participant's complete second-session recording without retraining. This yielded 73.2\% (SD\,=\,6.8) macro and 70.4\% (SD\,=\,7.2) micro window-level accuracy. Leave-one-user-out window-level evaluation yielded 63.1\% (SD\,=\,7.5) and 55.3\% (SD\,=\,8.1), respectively. These transfer results remain above six-class chance but are substantially below within-session LORO. Re-donning changes contact location, pressure, and impedance, while anatomical variation changes the planar current paths; person-independent transfer is therefore the hardest evaluated regime. The gap bounds the current system's practical generalization rather than supporting a deployment-ready claim.

\subsection{Protocol Ablation}
\label{sec:ablation_cross}

All 12 participants completed a matched comparison under a standard 40-channel adjacent-injection protocol on the same device and pipeline, which reached window-level LORO accuracy of 76.3\% (SD\,=\,3.4) for macro- and 79.7\% (SD\,=\,5.2) for micro-gestures. Multi-depth scanning improved on this by 15.1 and 10.4 percentage points despite using fewer channels per frame (35 vs.\ 40), and every participant improved. This supports multi-depth, multi-path scanning as a stronger starting point for the planar watch back, though it does not establish optimality across planar EIT configurations.
\subsection{Latency and Power}
\label{sec:latency_power}

To approximate a more realistic on-watch deployment, we separately built an online dual-core pipeline that runs entirely on the ESP32-S3: one core handles acquisition and rolling-window updates, and the other runs preprocessing and ExtraTrees inference. Predictions update in lockstep with the rolling window. The latency and power figures below are measured on this on-device path, not on the Wi-Fi logging path used in the user study. Frames arrive every 20.8\,ms at 48\,Hz. The measure-to-inference update takes 3.07\,ms for macro-gestures (1.01\,ms preprocessing and 2.06\,ms inference, SD\,=\,0.05\,ms) and 5.25\,ms for micro-gestures (0.75\,ms preprocessing and 4.50\,ms inference, SD\,=\,0.03\,ms), both fitting within one rolling-window update interval. End-to-end motion-to-photon latency averages 847.1\,ms for macro-gestures, dominated by the 0.83\,s feature window, and 72.9\,ms for micro-gestures. Running acquisition and inference continuously draws 35\,mA at 4.3\,V (0.15\,W), corresponding to approximately 8.6\,h from an idealized 300\,mAh, 4.3\,V battery. These are prototype measurements on the XIAO ESP32-S3, not an all-day smartwatch result.

\section{Limitations and Discussion}
\label{sec:discussion}

The central finding is that the planar watch-back patch carries gesture-discriminative EIT signal: within-session performance held across all 12 participants, multi-depth scanning beat adjacent injection for every participant, and transfer across sessions and users stayed above chance. The much lower transfer accuracies matter equally: EITWatch is a feasibility prototype, not a deployment-ready recognizer.

Its limitations set the agenda. We do not sweep strap force, watch rotation, contact impedance, or wrist tissue composition, so robustness is only indirectly probed. The prompted micro task relies on annotated intervals and a small vocabulary, leaving open-world spotting, false-activation rates, and live UI feedback for separate study. And while inference fits each 20.8\,ms frame, macro latency is dominated by its 0.83\,s window.

These point to clear next steps: integrate the front end into a watch case, improve transfer via don-time calibration or user-adaptive training, and broaden gesture sets and populations.

\bibliographystyle{ACM-Reference-Format}
\bibliography{references}

\end{document}